\documentclass[aip,preprint]{revtex4-1}
\usepackage{graphicx}
\usepackage{dcolumn}
\usepackage{bm}

\begin{document}

\preprint{AIP/123-QED}

\title{Effect of guide field on three dimensional electron shear flow
instabilities in collisionless magnetic reconnection\\
}

 \author{Neeraj Jain}
 \author{J\"{o}rg B\"{u}chner}%
\affiliation{Max Planck/Princeton Center  for Plasma Physics,  G\"{o}ttingen,
Germany}
 \affiliation{Max Planck Institute for Solar System Research,
Justus-Von-Liebig-Weg-3, G\"{o}ttingen, Germany}
%

\begin{abstract}
We examine the effect of an external guide field and current sheet
thickness on the growth rates and nature of three dimensional unstable modes of
an electron current sheet driven by electron shear flow. The growth rate of
the fastest growing mode drops rapidly with current sheet
thickness but increases slowly with the strength of the guide field. The fastest
growing mode is tearing type only for thin current sheets (half thickness
$\approx d_e$, where $d_e=c/\omega_{pe}$ is electron inertial length) and zero
guide field. For finite guide field or thicker current sheets, fastest growing
mode is non-tearing type. However growth rates of the fastest 2-D
tearing mode  and 3-D non-tearing mode are comparable for thin current sheets
($d_e < $half thickness $ < 2\,d_e$) and small guide field (of the order of the
asymptotic value of the component of magnetic field supporting electron
current sheet).  It is shown that
the general mode resonance conditions for electron-magnetohydrodynamic (EMHD)
and
magnetohydrodynamic (MHD) tearing
modes depend on the effective dissipation mechanism (electron inertia and
resistivity in
cases of EMHD and MHD, respectively). The usual tearing mode resonance
condition ($\mathbf{k}.\mathbf{B}_0=0$, $\mathbf{k}$
is the wave vector and $\mathbf{B}_0$ is equilibrium magnetic field) can be
recovered from the general resonance conditions in the limit of weak
dissipation.
Necessary
conditions (relating current sheet thickness, strength of the guide field and
wave numbers) for the existence of tearing mode are obtained from the general
mode resonance conditions.

\end{abstract}

\maketitle

\section{\label{sec:intro}Introduction}
Magnetic reconnection is a plasma process in which topological changes of
magnetic field lines release the energy stored in magnetic field in the form of
kinetic energy and heat. 
It is considered to be the cause of  the release of
magnetic energy in solar flares, sub-storms in Earth's magnetosphere, sawtooth
crashes in tokamaks and many astrophysical systems, e.g., accretion disk.
The topological changes of magnetic field lines take place in a current sheet
and require dissipation in the current sheet. 
In the absence of collisions, lack of dissipation allows the current sheet to
thin down to microscopic scales, such as, electron and ion inertial lengths,
and an effective dissipation is provided by micro-physical plasma processes. As
a result,
the reconnecting current sheet develops a two scale structure (along its
thickness), viz., an electron
current sheet with thickness of the order of an electron inertial length,
$d_e=c/\omega_{pe}$ embedded inside an ion current sheet  with thickness of the
order of an ion inertial length, $d_i=c/\omega_{pi}$. Reconnection of field
lines takes place in electron current sheets and couples to ion and then further
to very large MHD scales. 
 
The electron and ion current sheets are susceptible to a
variety of
instabilities. These instabilities have potential to affect the
rate and structure of reconnection. The role of ion scale 
instabilities in reconnection has been discussed in a review by B\"{u}chner
and Daughton (2006)\cite{buechner2006}. On electron scales,  
the electron current sheet (ECS) can be unstable to electro-magnetic
instabilities driven by the gradients in the ECS or to electrostatic Buneman
\cite{buneman1958} instability driven by relative drift of electrons and ions.
The electrostatic Buneman instability grows
(typical growth rates $\sim$ fraction of $\omega_{pe}$) much faster than  the
electro-magnetic instabilities without affecting magnetic fields.  
Growth of the electro-magnetic tearing and non-tearing instabilities leads to
the generation  of flux ropes/plasmoids \cite{daughton2011,markidis2013} and
filamentation of the ECS \cite{che2011}, respectively. The
conditions under
which an ECS filaments or generates flux ropes/plasmoids depend on the presence
of an external guide
field and thickness of the current sheet.
Recently it was shown that, depending upon thickness of the ECS, 
Electron Shear Flow Instabilities (ESFI) can grow both as
3-D oblique tearing
modes which generate flux ropes/plasmoids 
and/or non-tearing modes which filament the ECS \cite{jain2014a,jain2014b}.
Without an external guide field, the tearing mode instability dominates over
non-tearing modes only if the sheet is thin, with half thickness 
close to an electron
inertial length \cite{jain2014a,jain2014b}. Otherwise non-tearing mode
instabilities dominate.

In this paper, we perform linear stability analysis for ESFI of an ECS in the
presence of an external guide field.  We employ an
electron-magnetohydrodynamic
(EMHD) \cite{kingsep90} model. 
 Studies of electron shear flow instabilities in EMHD approximation
have earlier been reported but without guide field
\cite{basova1991,das01,drake94,jain03,jain04,jain2012b,gaur2012}.  These
instabilities have
also been referred as current driven sausage and kink instabilities, because, in
EMHD electron
flow is equivalent to current \cite{das01,jain03,jain04}.

In the next section we briefly describe electron-magnetohydrodynamic
model and obtain the linearized EMHD equations for ESFI.
Section \ref{sec:results} discuss effects of guide field on ESFI. Necessary
conditions for the existence of EMHD tearing mode are obtained in Section
\ref{sec:resonance}.
We conclude this paper in
section \ref{sec:conclusion}.

\section{\label{sec:emhd}Electron-MHD model}
Electron-magnetohydrodynamic  (EMHD) model is a fluid model for  
electron dynamics in a stationary background of ions. It is valid
for spatial scales smaller than $d_i$ and time scales smaller than
$\omega_{ci}^{-1}$. In EMHD, electron dynamics is described by electron momentum equation coupled with Maxwell's equations. An evolution equation for magnetic field $\mathbf{B}$ can be obtained by eliminating 
electric field from the electron momentum equation using Faraday's law \cite{kingsep90}.
\begin{eqnarray}
\frac{\partial}{\partial t}(\mathbf{B}-d_e^2\nabla^2\mathbf{B})&=&\nabla \times
[\mathbf{v}_e\times (\mathbf{B}-d_e^2\nabla^2\mathbf{B})]\label{eq:emhd1},
\end{eqnarray}
where, $\mathbf{v}_e=-(\nabla\times\mathbf{B})/\mu_0n_0e$ is electron fluid velocity. In addition to
ignoring the ion dynamics, Eq. (\ref{eq:emhd1}) assumes uniform electron
number
density $n_0$ and incompressibility of the electron fluid. The displacement
current is ignored under the assumption $\omega << \omega_{pe}^2/\omega_{ce}$.
In EMHD, the frozen-in condition of magnetic field breaks down due to the
electron inertia (which is contained in the definition of $d_e \propto
\sqrt{m_e}$). In the absence of electron
inertia ($d_e\rightarrow 0$), Eq. (\ref{eq:emhd1}) represents the condition that magnetic field
is frozen in
the electron fluid. 

The equilibrium magnetic field is taken to be
$\mathbf{B}_{0}=B_{y0}\tanh(x/L)\hat{y}+B_{z0}\hat{z}$  corresponding to a
current density
$\mathbf{J}_{0}=(B_{y0}/\mu_0L)\mathrm{sech}^2(x/L)\hat{z}$, where $L$ is the
half thickness of the electron current sheet. 
 For stationary ions,
electron fluid velocity is related to current density by the relation 
$\mathbf{J}=-n_0e\mathbf{v}_{e}$.  In the limit of cold electrons, bipolar
electrostatic electric field co-located with the electron current sheet balances
the Lorentz force in the current sheet. Small deviations from charge neutrality
in the electron current sheet can support the bipolar electric field
\cite{li2008}. This force balance is different from the force balance between
pressure gradient and Lorentz force as in the case of a Harris current sheet.
The bipolar electrostatic electric field and the new force balance in electron
current sheet have been observed in particle-in-cell
simulations\cite{li2008,chen2011}, laboratory experiments \cite{yoo2013},  and
space observations \cite{chen2009}. EMHD equations linearized about this
equilibrium can be written as,

\begin{eqnarray}
d_e^2\frac{d^2v_x}{dx^2}-(1+k^2d_e^2)v_x&=&
 -\frac{d_e^2\omega_{ce}^2}{B_{y0}^2}\frac{F-d_e^2F''}{\bar{\omega}}b_x
-\frac{d_e^2\omega_{ce}^2}{B_{y0}^2}\frac{F(F-d_e^2F'')}{\bar{\omega
} ^2 } v_x \nonumber\\
&&+ \frac{k_z(v_0-d_e^2v_0'')}{\bar{\omega}}v_x\label{eq:linemhd1}\\
d_e^2\frac{d^2b_x}{dx^2}-(1+k^2d_e^2)b_x&=&\frac{F-d_e^2F''}{\bar{
\omega}}v_x\label{eq:linemhd2}
\end{eqnarray}
where $\bar{\omega}=\omega-k_zv_0$, $k^2=k_y^2+k_z^2$, $\omega_{ce}=eB_{y0}/m_e$
and 
$F=\mathbf{k}.\mathbf{B}_0$. 
The perturbed x-components of electron flow velocity ( $v_x$ ) and magnetic
field ( $b_x$ ) are Fourier transformed in $y$, $z$, and $t$. The sub-script 'e'
is dropped from electron flow velocity and a prime ($\prime$) over an
equilibrium variable denotes derivative with respect to x. Equations
(\ref{eq:linemhd1}) and (\ref{eq:linemhd2}) are solved for confined eigen
functions $v_x$ and $b_x$ corresponding to eigen frequency $\omega$ which is
related to $k_y$ , $k_z$ , $L$ and $B_{z0}$ by a dispersion relation. Results
are presented in normalized variables. The magnetic field is normalized by
$B_{y0}$ , length by the electron inertial length $d_e$ , time by the inverse
electron cyclotron frequency $\omega_{ce}^{-1}=(eB_{y0}/m_e)^{-1}$, and velocity
by the electron Alfven velocity $v_{Ae}=d_e\omega_{ce}$ . Under this
normalization, $\mathbf{J} = -\mathbf{v}_e$ holds. 
\begin{figure*}
 \includegraphics[width=\textwidth,height=0.5\textheight]
 {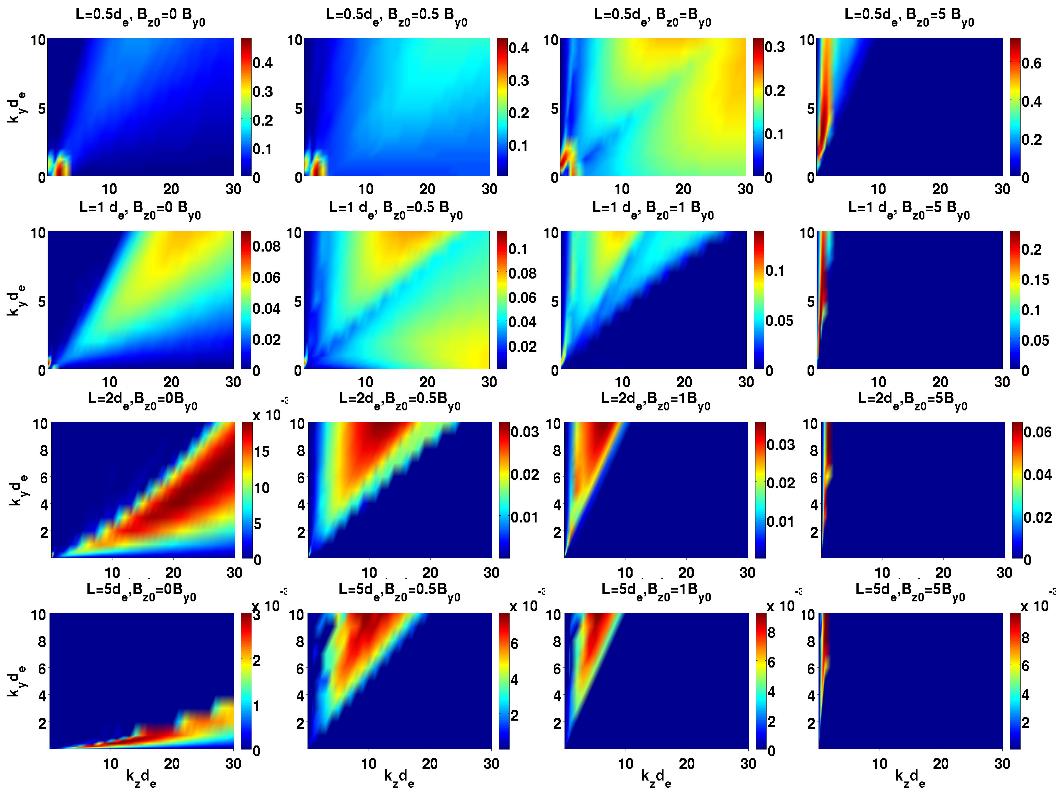}
 \caption{\label{fig:g_kykz} Growth rates $\gamma/\omega_{ce}$ (color) of ESFI
in $k_y$-$k_z$ space for $L/d_e=0.5$, 1.0, 2.0 and 5.0 (along column). For each
value of $L$, $B_{z0}/B_{y0}=0$, 0.5, 1.0 and 5.0. (along row).}
\end{figure*}
\section{Tearing and non-tearing electron shear flow
instabilities\label{sec:results}}
We define an unstable mode to be tearing mode if (1) $v_x$ is an odd function
of $x$ at
some position $x=x_0$ (inside the ECS) so that $v_x(x_0)=0$
and (2) $b_x(x_0)\neq 0$. All other modes are considered non-tearing type. Note
that the free energy source for all the unstable modes (both tearing and
non-tearing) is electron shear flow which is equivalent to electric current for
stationary ions.

 Growth
rates in $k_y$-$k_z$ space for various values of $L$ and $B_{z0}$  are shown in
Fig. \ref{fig:g_kykz}.  For a fixed
value of current sheet
thickness, the domain of unstable modes shift to smaller values of $k_z$.
The growth rates $L>d_e$ are significantly  affected even for small guide
field. On the other hand growth rates
for $L<d_e$ are not affected much when strength of the guide field is small.
This behavior is expected as Lorentz force terms in electron momentum
equation are negligible in the limit $L<<d_e$. It can also be shown that Eqs.
(\ref{eq:linemhd1}) and (\ref{eq:linemhd2}) becomes independent of guide field
in this limit.  
In the limit of $L<<d_e$, $F-d_e^2F'' \approx -d_e^2k_yB_{y0}/L^2$is
independent of  guide field. Therefore guide field appears only in the
expression for $F=k_yB_{y0}+k_zB_{z0}$ in the second
last term on the RHS of Eq. (\ref{eq:linemhd1}). In the limit 
$L<<d_e$, this term can be neglected  in comparison to the last
term on RHS of Eq. \ref{eq:linemhd1}, thus making Eqs. (\ref{eq:linemhd1})
and (\ref{eq:linemhd2}) independent of the guide field.

\begin{figure}
 \includegraphics[width=0.5\textwidth,height=0.4\textheight]
 {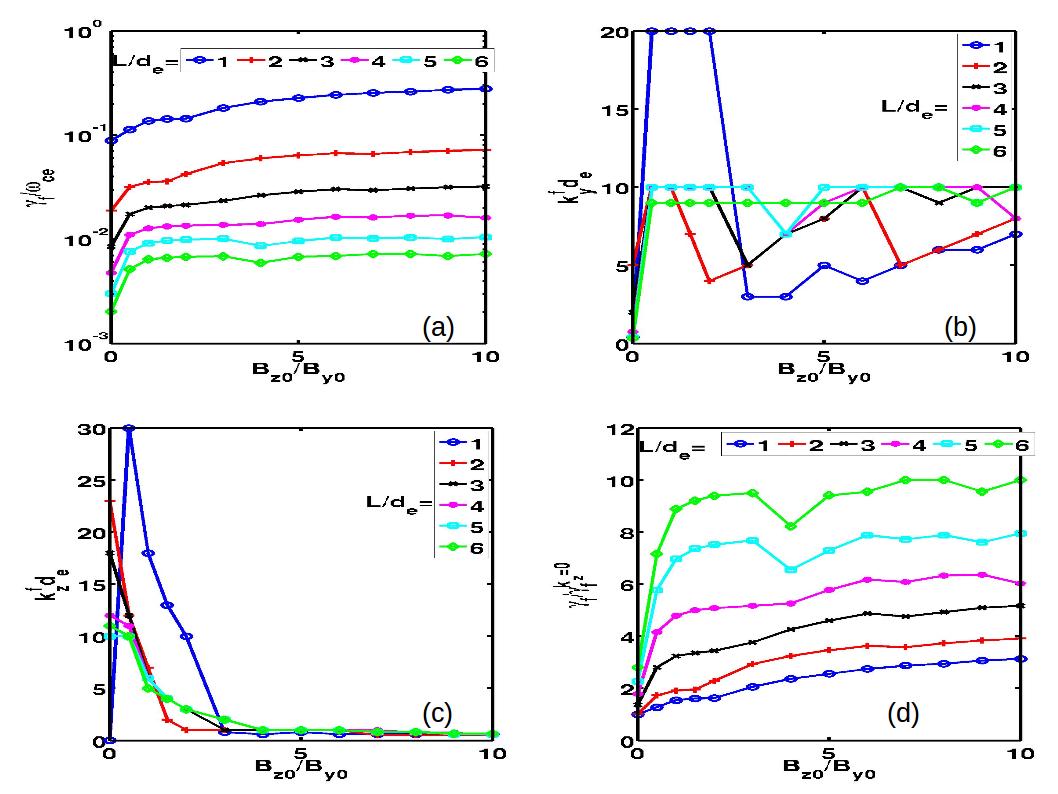}
 \caption{\label{fig:fastest_mode} Growth rates $\gamma_f$ (a), wave numbers
$k_y^f$ (b) and $k_z^f$ (c) of the fastest growing mode as functions of guide
field $B_{z0}$ for various current sheet thicknesses. In (d), 
ratio $\gamma_f/\gamma_f^{k_z=0}$, where $\gamma_f^{k_z=0}$ is 
the growth rate of the fastest 2-D ($k_z=0$)
tearing mode, as a function of the guide field.}
\end{figure} 

The maximum growth rate drops rapidly with $L$ for a given
value of the guide field.  The maximum growth
rate increases slowly as the guide field is increased
for fixed current sheet thickness except for $L=0.5\,d_e$ (in Fig.
\ref{fig:g_kykz}) for which maximum growth
rate first drops and then increases. An electron shear
flow instability whose 
growth rate increases  with the strength of the guide field was also
reported in 3-D
particle-in-cell simulations of electron current sheet \cite{che2011}.
The variation of the growth rate ($\gamma_f$) of the fastest growing mode
(maximum growth rate in
$k_y$-$k_z$ space) with the strength of the guide field for various values of
$L$ is shown in Fig. \ref{fig:fastest_mode}a. The rate of
increase of the growth rate is faster for weak
strength of the guide field ($B_{z0}/B_{y0} < 1$) as compared to the rate for
large guide field ($B_{z0}/B_{y0} > 5$). In between the fast and
slow rise of the growth rate with guide field, a plateau forms.  The rate of the
fast rise  of $\gamma_f$ for weak guide field is higher for large
values of $L$ while opposite is true for the slow increase for strong guide
field. The range of the guide field values for the plateau  depends
on $L$. 

The wave numbers, $k_y^f$ and $k_z^f$, of the fastest growing mode are shown in
Figs. \ref{fig:fastest_mode}b and \ref{fig:fastest_mode}c. The wavenumber
$k_y^f$ takes a jump to large values as soon as the guide field becomes finite
and remains at this value for a range of the values of guide field. On further
increasing the guide field, $k_y^f$ drops to rise again. This trend of
variation of $k_y^f$ with the guide field can be seen for all values of $L$. The
jump in the value of $k_y^f$ is the largest for $L=d_e$. The wave number $k_z^f$
has large jump as soon as guide field becomes finite only for $L=d_e$. After the
initial jump $k_z^f$ drops. For $L>d_e$, $k_z^f$ drops with increasing guide
field. For all values of $L$, $k_z^f$ saturates at $k_z^fd_e\approx 1$ for large
values of guide field. The sudden changes in $\gamma_f$, $k_y^f$ and $k_z^f$ as
soon as guide field becomes finite indicate that the inclusion of guide field
triggers a new kind of unstable mode different from tearing mode.

\begin{figure*}
 \includegraphics[width=\textwidth,height=0.5\textheight]
 {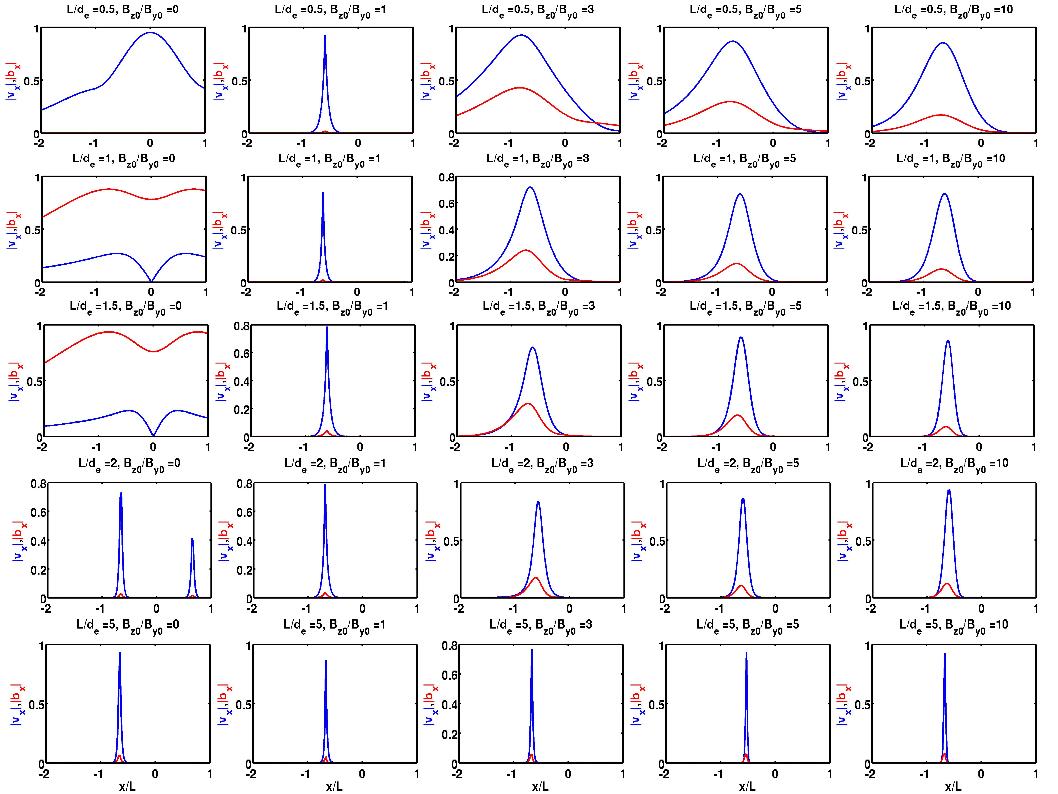}
 \caption{\label{fig:vxbx_forgmax} Absolute values of complex eigen functions
$v_x$ (blue) and $b_x$ (red) corresponding to the fastest growing mode for
$L/d_e=0.5, 1, 1.5, 2$ and 5 (varying along a column), and $B_{z0}/B_{y0}=0, 1,
3, 5$ and 10 (varying along a row). Horizontal axis is in units of $L$ and
shows only a small region in $x$.}
\end{figure*}

In order to distinguish between tearing and non-tearing modes, we plot absolute
values of complex eigen functions $v_x$ and $b_x$ of the fastest growing mode in
Fig. \ref{fig:vxbx_forgmax}. For tearing mode, x-component of magnetic field
($B_x$) and velocity ($V_x$) in real space should be finite and zero,
respectively, at the mode rational surface. These components can be expressed
as $B_x=\mathrm{Re}[|b_x|\exp(i\theta_b+ik_y y+ik_zz-i\omega t)]$ and
$V_x=\mathrm{Re}[|v_x|\exp(i\theta_v+ik_y y+ik_zz-i\omega t)]$, where
$\tan(\theta_v)=\mathrm{Im}(v_x)/\mathrm{Re}(v_x)$ and
$\tan(\theta_b)=\mathrm{Im}(b_x)/\mathrm{Re}(b_x)$. In Fig.
\ref{fig:vxbx_forgmax}, $|v_x|=0$ and $|b_x|\ne0$ (and therefore so for $V_x$
and $B_x$) only for $L/d_e=1$ and 1.5 and $B_{z0}=0$. Therefore tearing mode is
the fastest growing mode in a very small range of $L-B_{z0}$ parameter space.  

Although the fastest growing mode is non-tearing in the presence of finite
guide field, 2-D tearing mode ($k_z=0$) is not affected by the guide field
because guide field disappears from eigen value Eqs. (\ref{eq:linemhd1}) and
(\ref{eq:linemhd2}) for $k_z=0$.
 For $L=d_e$ and $B_{z0}=B_{y0}$, growth rate of the fastest 2-D tearing mode 
 ($k_yd_e=0.4$)  is comparable to that of the fastest
growing 3-D non-tearing mode (Fig. \ref{fig:g_kykz}). In order to determine
the
relative importance of the
tearing and non-tearing modes as a function of $L$ and $B_{z0}$, we compare the
growth rate of the fastest 3-D mode ($\gamma_f$) and that of the fastest 2-D
($k_z=0$) tearing mode ($\gamma_f^{k_z=0}$) in Fig. \ref{fig:fastest_mode}d.
The ratio $\gamma_f/\gamma_f^{k_z=0}$ increases with both the current sheet
thickness and the strength of the guide field. For zero guide field, the ratio
$\gamma_f/\gamma_f^{k_z=0}$ is of the order of unity (varies from 1-3 for
$L/d_e=$1-6). The ratio increases with the strength of the guide field but
remain order of unity for thinner current sheets ($L/d_e=1$ and 2) for guide
field as large as  $B_{z0}/B_{y0}=10$. The initial rate of increase of the
ratio with guide field is larger for thicker current sheets. Therefore the
ratio increases to much larger values even for $L>2\,d_e$. It is expected that
tearing mode can grow simultaneously with the non-tearing modes for thin
current sheets and small value of guide field.


\section{Necessary conditions for EMHD tearing mode\label{sec:resonance}}
Applying the tearing mode conditions, viz., $v_x(x_0)=0$ and $b_x(x_0)\neq 0$,
in  Eqs. (\ref{eq:linemhd1}) and
(\ref{eq:linemhd2}), we get necessary conditions for the existence of tearing
eigen
functions.
\begin{eqnarray}
[\mathbf{k}.\mathbf{B}_0-\mathbf{k}.\mathbf{B}_0'']_{x=x_0}&=&0,
\label{eq:kdotb}\\
\left[\frac{1}{b_x}\frac{d^2b_x}{dx^2}\right]_{x=x_0}&=&1+k^2, \label{eq:bx_x0}
\end{eqnarray}
where we have used $[d^2v_x/dx^2]_{x=x_0}=0$ as $v_x$ is an odd function about
$x=x_0$. Eqs. (\ref{eq:kdotb}) and (\ref{eq:bx_x0}) are  
only necessary conditions for the
existence of tearing type eigen functions which could either be stable or
unstable.  Since RHS of Eq. \ref{eq:bx_x0} is always finite, $b_x(x_0)$ and 
$[d^2b_x/dx^2]_{x_0}$ can not be zero, except when both $b_x(x_0)$ and
$[d^2b_x/dx^2]_{x_0}$ approach zero simultaneously.

Eq. (\ref{eq:kdotb}) is the mode resonance condition for EMHD tearing
mode. Note that this condition is different from
 the usual resonance condition ($\mathbf{k}.\mathbf{B}_0=0$) for tearing
mode. It reduces to $\mathbf{k}.\mathbf{B}_0=0$ only for equilibrium scale
length much larger than electron inertial length. In fact mode resonance
condition for MHD tearing mode should also be different from
$\mathbf{k}.\mathbf{B}_0=0$. This can be seen by applying conditions for
tearing mode eigen functions in Eq. (14) of Furth et al. \cite{furth1963}
Assuming uniform density, we get for MHD
\begin{eqnarray}
 \mathbf{k}.(\mathbf{B}_0-L_{\eta}^2\mathbf{B}_0'')&=&0\label{eq:kdotb_mhd}
\end{eqnarray}
where $L_{\eta}=\eta/4\pi\omega$ is the resistive scale length and $\eta$ is
the resistivity. Eq. \ref{eq:kdotb_mhd} reduces to $\mathbf{k}.\mathbf{B}_0=0$
only when equilibrium scale length is much larger than the resistive scale
length.

Now we obtain some conditions on the existence of EMHD tearing mode from
mode resonance condition Eq. (\ref{eq:kdotb}). Substituting for the equilibrium
magnetic field 
$\mathbf{B}_0$ in Eq. (\ref{eq:kdotb}), the values of $x_0$ can be obtained
from following equation.
\begin{eqnarray}
\tanh^3\left(\frac{x_0}{L}\right)-\left(1+\frac{L^2}{2}\right)\tanh\left(\frac{
x_0}{L}\right)-\frac{k_zB_{z0}L^2}{2k_y}&=&0\label{eq:x0}
\end{eqnarray}

\begin{figure}
 \includegraphics[width=0.5\textwidth,height=0.4\textheight]
 {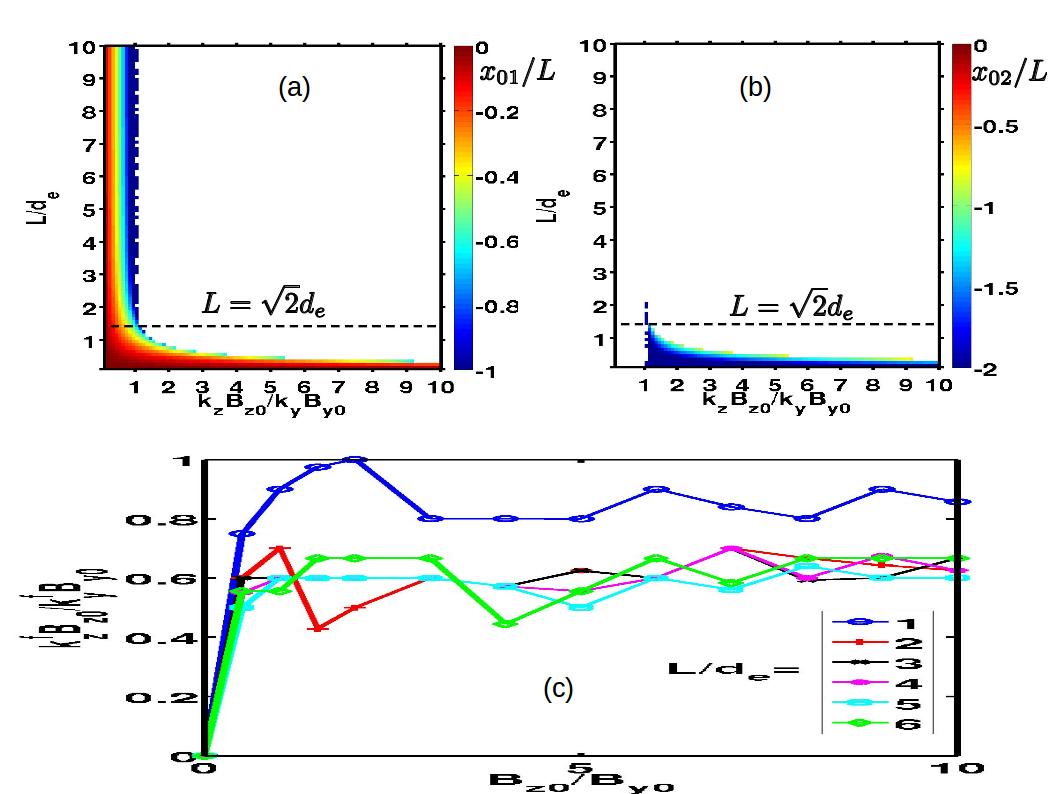}
 \caption{\label{fig:x0_formula} Real roots $x_{01}/L$ (a) and $x_{02}/L$
(b) of Eq. (\ref{eq:x0}). Color
scales are saturated at $x_{01}/L=-1$ and $x_{02}/L=-2$. No real root exists in
the white regions in $L$-$k_zB_{z0}/k_y$ space. In (c), quantity
$k_z^fB_{z0}/k_y^fB_{y0}$ for the fastest growing modes obtained from the
solutions of Eqs. (\ref{eq:linemhd1}) and (\ref{eq:linemhd2}) is plotted as a
function of guide field for various values of current sheet thickness.} 
\end{figure}

For zero guide field, $B_{z0}=0$, only real solution of
Eq. (\ref{eq:x0}) is $x_0=0$, as is expected. For finite guide field, $x_0$
can have non-zero values except for $k_z=0$. One of the three roots of Eq.
(\ref{eq:x0}) is complex everywhere in the $L$-$k_zB_{z0}/k_y$ parameter
space. The other two roots $x_{01}$ and $x_{02}$ have real values in
regions of the $L$-$k_zB_{z0}/k_y$ parameter space and are plotted in Fig.
\ref{fig:x0_formula}. There exist no real solution in the parameter space
$L>\sqrt{2}d_e$ and $k_zB_{z0}/k_y > 1$. In the limit $L>>\sqrt{2}d_e$ and
$k_zB_{z0}/k_y >> 1$, Eq. (\ref{eq:x0}) reduces to
$\tanh^3(x_0/L)=k_zB_{z0}L^2/2k_y$ which has no real solution because
$\tanh(x_0/L) \le 1$. 
When $k_zB_{z0}/k_y<1$, real solutions of $x_0$ exist for
any value of $L$. In this limit, value of $x_{01}/L$ does not depend on  $L$ for
$L>>\sqrt{2}d_e$ but does depend on $k_zB_{z0}/k_y$. For $L>>\sqrt{2} d_e$ and
$k_zB_{z0}/k_y<1$, Eq. (\ref{eq:x0}) can be simplified by neglecting the cubic
term to give $x_0/L=-\tanh^{-1}(k_zB_{z0}/k_y)$ which is independent of
$L$. This result can be derived from the condition $\mathbf{k}.\mathbf{B}_0=0$
which can be obtained from the resonance condition, Eq. (\ref{eq:kdotb}), in the
limit $L>>d_e$. In Fig. \ref{fig:x0_formula}, $x_{01}/L < 0.6$ for
$L<\sqrt{2}d_e$ and cubic term in Eq. \ref{eq:x0} can be neglected giving,
\begin{eqnarray}
 \tanh\left(\frac{x_{01}}{L}\right)&=&-\frac{k_zB_{z0}}{k_y}\frac{L^2/2}{L^2/2+1
} \label{eq:x01}.
\end{eqnarray}
Eq. (\ref{eq:x01}) can have real solutions for $x_{01}$ as long as the RHS $\le
1$ which is satisfied for $k_zB_{z0}/k_y \le 1$ for any $L < \sqrt{2}d_e$.
For $k_zB_{z0}/k_y > 1$, 
real $x_0$ exists only if $L$ is sufficiently small. For this reason,
range of the values of $L$ which allow real $x_{01}$ shrinks with increasing
$k_zB_{z0}/k_y > 1$. When $k_zB_{z0}/k_y\rightarrow 1$ and $L>\sqrt{2}d_e$,
$|x_{01}|/L \gg 1 $ pushing mode rational surface out of the electron current
sheet.  

The resonance condition, Eq. (\ref{eq:kdotb}), is only a necessary condition
for the existence of tearing eigen functions. Therefore it can not be used
with certainty to predict the existence of tearing instability. However, its
violation guarantees that tearing mode can not exist. The conditions for the
non-existence of tearing mode can be stated as follows.
\begin{eqnarray}
 \frac{k_zB_{z0}}{k_y}& >& 1,\,\, \mathrm{  if  } \,\, L >\sqrt{2} d_e\\
 &>&1+2/L^2,\,\, \mathrm{  if  } \,\, L < \sqrt{2} d_e
 \end{eqnarray}
Fig. \ref{fig:x0_formula}c shows that value of the quantity
$k_zB_{z0}/k_yB_{y0}$ calculated for the fastest growing remains smaller than
unity even for those value of  $L$ and $B_{z0}$ for which the fastest mode is
non-tearing. This is not in contrast to the conditions stated above as
$\mathbf{k}.(\mathbf{B}_0-\mathbf{B}_0'')=0$ is only a necessary condition for
the existence of tearing mode.

\section{\label{sec:conclusion}Conclusion and Discussion}
We have performed linear stability analysis of three dimensional electron shear
flow instabilities of an ECS.  The unstable domain of wave numbers shifts
towards
smaller values of $k_z$ with the increasing strength of the guide field. The
wave number $k_z^f$ (along the direction of guide field)  of the fastest growing
mode drops with guide field and saturates at $k_z^fd_e\approx1$ for large guide
field. The growth rate of the fastest mode drops rapidly with current sheet
thickness but increases slowly with the strength of the guide field. On
examining the eigen functions of unstable modes we found that the fastest
growing mode is no longer tearing mode for finite guide field even for thin
current sheets ($L\approx d_e$). However growth rate of the fastest 2-D
($k_z=0$) tearing mode  is comparable to the growth rate of the fastest mode
for thin current sheets ($1<L/d_e<2$) for small guide field. Therefore, in the
nonlinear evolution of thin current sheets with a small guide field both
tearing and non-tearing modes are expected to grow. For thicker current sheets
the growth rate of the fastest 2-D tearing mode is much smaller than the
fastest mode and thus the evolution is expected to be dominated by non-tearing
mode. However, in cases where reconnection is driven from boundaries, the
evolution may be dominated by tearing mode even for large guide field and
thicker current sheets. 

A general mode resonance condition,
$\mathbf{k}.({\mathbf{B}_0-\mathbf{B}_0''})$, for EMHD tearing mode was
obtained. It was shown that the general resonance condition for MHD tearing
mode, $\mathbf{k}.{\mathbf{B}_0-L_{\eta}^2\mathbf{B}_0''}$, is different from
the usual resonance condition $\mathbf{k}.\mathbf{B}_0=0$. However the general
resonance condition for EMHD (MHD) tearing mode reduces to
$\mathbf{k}.\mathbf{B}_0=0$ if equilibrium scale length is much larger than the
electron inertial length (resistive scale length). Necessary conditions for
existence of tearing mode were obtained.
 These conditions relate wave
numbers, current sheet thickness and strength of the guide
field. 
\begin{acknowledgments}
This work was supported by Max-Planck/Princeton Center for Plasma Physics at the 
Max Planck Institute for Solar System Research, Justus-von-Liebig-Weg-3,
G\"{o}ttingen, Germany. We thank Dr. Bernhardt
Bandow for his help to numerically optimize the EMHD code. 
\end{acknowledgments}


%

\end{document}